\documentclass[aps,prb,preprint,groupedaddress,showpacs,floatfix]{revtex4-1}

\usepackage{graphics}
\usepackage{epsfig}

\begin{document}
\title{Effects of rare earth ion size on the stability of the coherent Jahn-Teller distortions
in undoped perovskite manganites}
\author{T. F. Seman,$^1$ K. H. Ahn,$^1$ T. Lookman,$^2$ A. Saxena,$^2$ A. R. Bishop,$^2$ and P. B. Littlewood$^3$}
\affiliation{
$^1$Department of Physics \\ New Jersey Institute of Technology
\\ Newark, New Jersey 07102, USA \\
$^2$Theoretical Division \\
Los Alamos National Laboratory \\ Los Alamos, New Mexico 87545, USA \\
$^3$Physical Sciences and Engineering Division \\ Argonne National Laboratory \\
 Argonne, Illinois 60439, USA
}


\begin{abstract}
    We present a theoretical study on the relation between the size of the rare earth ions,
    often known as chemical pressure, and the stability of the coherent Jahn-Teller distortions
    in undoped perovskite manganites. Using a Keating model expressed in terms of atomic scale
    symmetry modes, we show that there exists a coupling between the uniform shear distortion and the staggered
    buckling distortion within the Jahn-Teller energy term. It is found that this coupling
    provides a mechanism by which the coherent Jahn-Teller distortion is more stabilized by
    smaller rare earth ions. We analyze the appearance of the uniform shear distortion
    below the Jahn-Teller ordering temperature,
    estimate the Jahn-Teller ordering temperature and its variation
    between NdMnO$_3$ and LaMnO$_3$, and obtain the relations between distortions.
    We find good agreement between theoretical results and experimental data.
\end{abstract}


\maketitle


\section{Introduction}
Since the discovery of the colossal magnetoresistance effect, a lot of attention has
focused on a class of materials known as perovskite
manganites.~\cite{Helmolt93,Jin94,Millis95,Millis96,Roder96,Salamon01}
These materials have the chemical formula in the form of $RE_{1-x}AK_x$MnO$_3$, where
$RE$ and $AK$ represent the rare earth and alkali metal elements, and have a perovskite
structure. One of the major research themes for these materials is the relation between
their physical properties and the average size of ions at the $RE$/$AK$ site, often known
as the chemical pressure effect. The size of the $RE$/$AK$ ion is usually parameterized by
a {\it tolerance factor} and one of the most important phase diagrams for these materials
has been the one in the temperature versus tolerance factor plane for a fixed
30\% ($x=0.3$) doping ratio.~\cite{Hwang95} The $RE$/$AK$ ions with size smaller than
the space created by the surrounding MnO$_6$ octahedra induce buckling of the Mn-O-Mn
bonds, observed through various structural
refinement analyses.

How this buckling distortion affects the properties of
manganites has been controversial.
It is well known from experimental observations that there is  strong competition
between the insulating phase with a coherent Jahn-Teller (JT) distortion and the metallic
phase without such distortion.~\cite{Salamon01}
So far, most attention has centered on
the impact of the buckling on the metallic phase, in particular, the
possible change in  the effective Mn-O-Mn electron hopping parameter and the
band width.~\cite{Hwang95}
However, there has been a debate on whether the change of the hopping parameter
due to the  Mn-O-Mn bond angle change of several degrees
would be significant enough to explain the observed metal-insulator transition.~\cite{Dzero00,Fernandez-Baca98,Liu99,Lynn96,Radaelli97}
A less studied effect of the Mn-O-Mn bond buckling,
except for a few early efforts based on experimental data,~\cite{Louca01}
is the possibility that
the buckling distortion may significantly stabilize the insulating phase with a coherent
JT distortion.
The main goal of this paper is to examine such a possibility with
a simplified model of the perovskite manganites.
To be specific, we analyze the interplay
between the JT ordering and chemical pressure for undoped perovskite manganites.
The study on undoped manganites is merited  because they are not only parent compounds of doped
perovskite manganites, but also because one of the first multiferroic materials discovered
is an undoped manganite, TbMnO$_3$, with a relatively small rare earth element.~\cite{Kimura03Nature}
Therefore, the chemical pressure effect in undoped manganites reported in this paper would
also be relevant for  future studies on how the multiferroic property appears in $RE$MnO$_3$
with small $RE$ ions, as well as for the effect of chemical pressure on the distorted insulating
phase of doped manganites.

\section{Model system and energy expression}
\begin{figure}
    \leavevmode
    \epsfxsize4.5cm\epsfbox{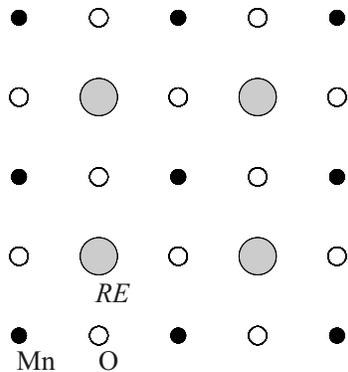}
    \caption{\label{fig:perov}
    Two-dimensional model for the perovskite structure considered in the text.
    }
\end{figure}

We study a two-dimensional (2D) model for the perovskite structure which
incorporates both  buckling and the JT distortions.
We define a 2D perovskite structure shown in Fig.~\ref{fig:perov} which includes the
following aspects of the 3D perovskite structure for undoped manganites:
(1) symmetry breaking distortion of oxygen ions around Mn ion, (2) chemical
pressure effect, which is the attraction of surrounding oxygen ions toward the small rare earth
ions, and (3) the rotation of oxygen ions with alternating directions
around Mn ions, which is, in effect, the buckling of
Mn-O-Mn bonds.

\begin{figure}
    \leavevmode \epsfxsize6.0cm\epsfbox{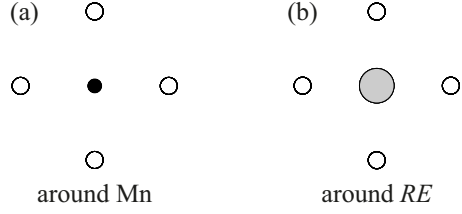}
    \caption{\label{fig:motif}
    Two structural motifs chosen for the 2D structure shown in Fig.~\ref{fig:perov}.
    }
\end{figure}

\begin{figure}
    \leavevmode \epsfxsize8.6cm\epsfbox{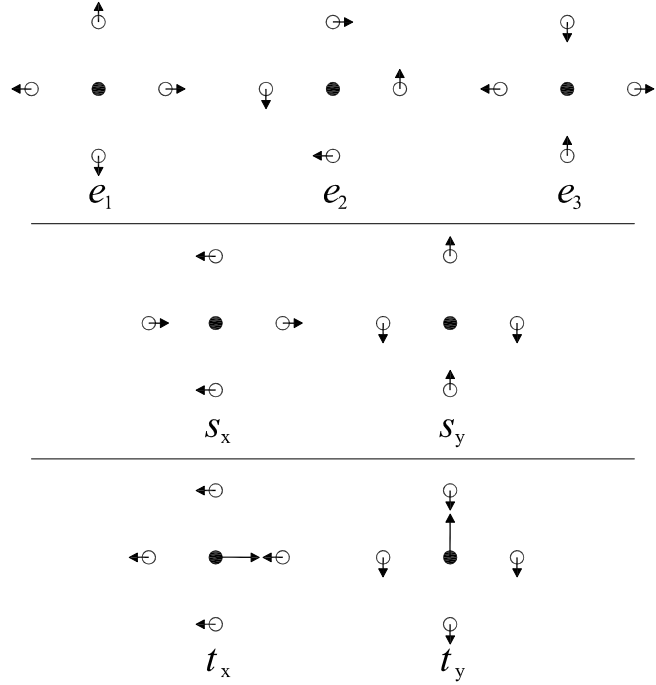}
    \caption{\label{fig:modes}
    Distortion modes for the motif around the Mn ion in Fig.~\ref{fig:motif}.
    Similar distortion modes, $e'_1$, $e'_2$, $e'_3$, $s'_x$, $s'_y$, $t'_x$, $t'_y$
    are defined for the motif around the $RE$ ion.
    }
\end{figure}

We apply the recently developed atomic scale description of lattice distortions~\cite{Ahn03,Ahn04}
to describe the elastic energy of the system. In this approach, atomic scale modes of lattice
distortions and their constraints are used instead of displacement variables. The structural
motifs can be chosen in any convenient way as long as they have the symmetry of the crystal
structure. We choose two ``structural motifs'', shown in Fig.~\ref{fig:motif}: one consists
of one Mn ion and four surrounding O ions and the other comprises of one
rare earth ($RE$) ion and four surrounding O ions. We obtain seven symmetry modes for each motif,
shown in Fig.~\ref{fig:modes} for the MnO$_4$ motif.~\cite{Norm} Similar symmetry modes are defined
for the $RE$O$_4$ motif and are distinguished with primes on the symbols in this paper.
The modes defined for each plaquette on the lattice are constrained by each other
because neighboring motifs share ions, which leads to constraint equations
between the Fourier components of the modes. In terms of these fourteen modes and constraint
equations, any distortion of the 2D perovskite structure shown in Fig.~\ref{fig:perov} can
be described.

\begin{figure}
    \leavevmode
    \epsfxsize8.6cm \epsfbox{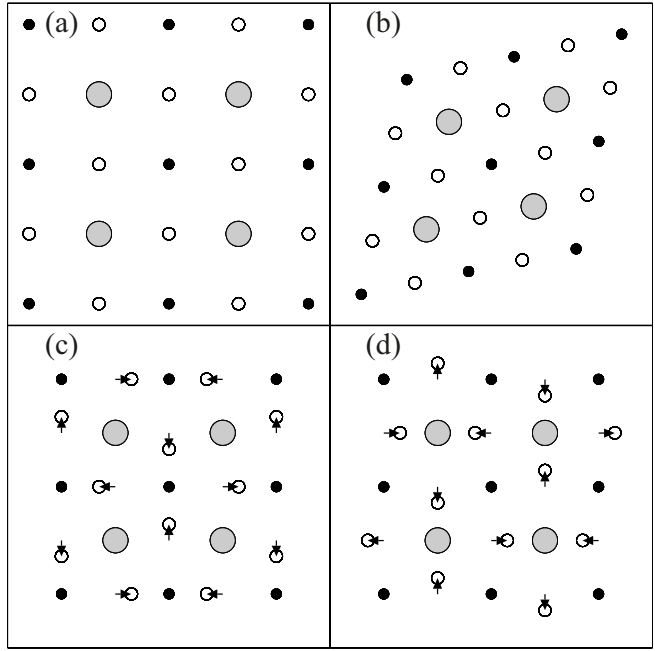}
    \caption{\label{fig:4modes}
    Four distortion modes considered in the current study: (a) uniform dilatation
    mode $e_{10}$, (b) uniform shear mode $e_{20}$, (c) staggered deviatoric mode $e_{3s}$,
    (d) staggered buckling mode $e'_{3s}$. All figures are drawn for the positive
    values of the modes with the Mn site at the left bottom corner chosen as the origin.
    }
\end{figure}

For the current study, since we are interested in the ordered state, we consider
distortions with wavevectors
$\vec{k}=(0,0)$ and $(\pi,\pi)$ only.
For these wavevectors, the constraint equations
are as follows, where we use subscripts 0 and $s$ to represent $\vec{k}$ = (0,0) and
$(\pi,\pi)$, respectively: $e_{10}=e'_{10}$, $e_{20}=e'_{20}$, $e_{30}=e'_{30}$, $s_{x0}=-s'_{x0}$,
$s_{y0}=-s'_{y0}$, $e_{1s}=-e'_{2s}$, $e'_{1s}=-e_{2s}$, $s_{xs}=s_{ys}=s'_{xs}=s'_{ys}=0$.
Rest of the modes are unconstrained, particularly, $e_{3s}$ and $e'_{3s}$.
We search for the interplay between the staggered deviatoric
distortion mode $e_{3s}$ and the staggered rotation
of O ions around Mn ion (or equivalently staggered Mn-O-Mn bond buckling mode) $e'_{3s}$,
where the latter is due to the
compression $e'_{10}=e_{10}$ by small rare earth ions. Therefore, we limit ourselves to
the modes $e'_{10}=e_{10}$, $e_{20}=e'_{20}$, $e_{3s}$, and $e'_{3s}$, shown in Fig.~\ref{fig:4modes}.
We include the uniform shear mode $e_{20}=e'_{20}$ because it is coupled to $e'_{3s}$ through the
JT term, as will be discussed later in this paper.

\begin{figure}
    \leavevmode
    \epsfxsize6.5cm \epsfbox{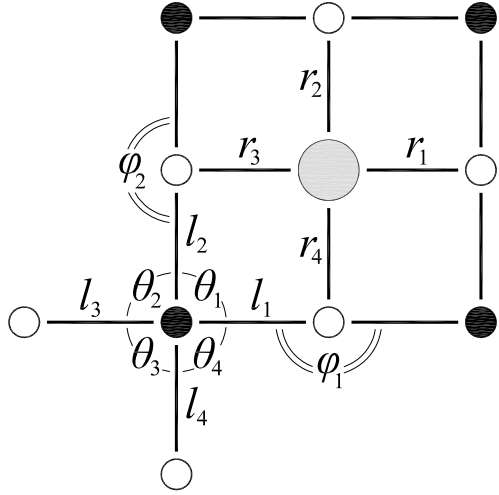}
    \caption{\label{fig:Keating}
    The Keating variables considered for each Mn ion. $l_1$, $l_2$, $l_3$, and $l_4$
    represent the Mn-O bond lengths. $\theta_1$, $\theta_2$, $\theta_3$, and $\theta_4$
    represent the O-Mn-O bond angles. $r_1$, $r_2$, $r_3$, and $r_4$ represent $RE$-O
    bond lengths. $\varphi_1$ and $\varphi_2$ indicate Mn-O-Mn bond angles.
    }
\end{figure}

Even though it is possible to analyze an energy expression including higher order
symmetry-allowed anharmonic energy terms, such a method would generate many parameters
and would make the model less predictive. Therefore, we start with a Keating model with
a small number of parameters,~\cite{Keating66,Littlewood86} and map the Keating model onto the approach
based on the symmetry modes. In the Keating approach, the elastic energy is represented in terms of
bond length and bond angle changes from equilibrium. For our 2D perovskite structure,
we consider the following set of Keating variables and harmonic moduli for each Mn ion, as shown in
Fig.~\ref{fig:Keating}: $\delta l_n$ ($n$ = 1, 2, 3, 4) and modulus $a_1$ for Mn-O bond length change,
$\delta \theta_n$ ($n$ = 1, 2, 3, 4) and $b_1/4$ for $90^{\circ}$
O-Mn-O bond angle change, $\delta r_n$ ($n$ = 1, 2, 3, 4) and $a_2$ for
$RE$-O bond length change, and $\delta \varphi_n$ ($n$ = 1, 2)
and $b_2/4$ for $180^{\circ}$ Mn-O-Mn bond angle change.
We note
that the MnO$_4$ motif is considered as relatively stiff compared to other components of the
structure, so that $a_1 \gg a_2$ and $b_1 \gg b_2$.

We consider the following Keating elastic energy expression per Mn ion,
\begin{eqnarray}
    E_{\rm elastic} &=& \frac{1}{2} a_1 \sum_{n=1,2,3,4} (\delta l_{n})^2 +
    \frac{1}{2} b_1 \sum_{n=1,2,3,4} (\delta \theta_{n}/2)^2 \nonumber \\
    &+& \frac{1}{2} a_2 \sum_{n=1,2,3,4} (\delta r_{n})^2 + \frac{1}{2}
    b_2 \sum_{n=1,2} (\delta \varphi_{n}/2)^2.
\end{eqnarray}
We express the Keating variables
in terms of $e_{10}$, $e_{20}$, $e_{3s}$ and $e'_{3s}$. For example, we obtain
\begin{equation}
    \delta l_{1} = \frac{\sqrt{(1+e_{10}+e_{3s})^2+(e_{20}+e'_{3s})^2}-1}{2},
\end{equation}
\begin{equation}
    \delta \theta_{1} = \tan^{-1}\left(\frac{e_{20}+e'_{3s}}{1+e_{10}+e_{3s}}\right)
    +\tan^{-1}\left(\frac{e_{20}-e'_{3s}}{1+e_{10}-e_{3s}}\right).
\end{equation}
The Taylor expansion of $E_{\rm elastic}$ in terms of $e_{10}$, $e_{20}$, $e_{3s}$ and $e'_{3s}$
produces all the terms of any order. We make an approximation that $b_2$ is much smaller than
other parameters, as mentioned above, and drop the terms with $b_2$. We keep all harmonic
order terms and select the cubic and quartic order terms that are responsible for the Mn-O-Mn
bond buckling instability, which are shown below as $E_{\rm har}$, $E_{\rm cubic}$,
and $E_{\rm quartic}$.

We further define the JT energy per Mn ion $E_{\rm JT}$ and the
energy associated with the tolerance factor per Mn ion $E_{\rm tol}$ as follows:
\begin{eqnarray}
    E_{\rm JT} &=& -\frac{\lambda}{2}
    | \delta l_{1} + \delta l_{3} - \delta l_{2} -\delta l_{4} |, \\
    E_{\rm tol} &=& \frac{\tilde{p}}{2} (\delta r_{1} + \delta r_{2}
    + \delta r_{3} + \delta r_{4}),
\end{eqnarray}
where we define ``chemical pressure"
\begin{equation}
    \tilde{p}=C'_1 (1-t).
\end{equation}
The parameter $t$ is a two-dimensional analog of the tolerance factor for the 3D perovskite
structure, and
the coefficient $C'_1$ represents the coupling between
the average $RE$-O bond length and the tolerance factor $t$.
The ``chemical pressure" $\tilde{p}$ induces the shortening
of the average $RE$-O bond length due to small $RE$ ions.
We also define the JT distortion mode
\begin{equation}
e_{\rm JT} = (\delta l_{1} + \delta l_{3} - \delta l_{2} -\delta l_{4})/2,
\end{equation}
which represents the anisotropic bond length change, similar to
the 3D JT distortion modes, often written as $Q_2$ and $Q_3$ (Ref.~\onlinecite{Ahn01}).
The expression $E_{\rm JT} = -\lambda | e_{\rm JT} |$ is based on
the 3D JT energy $E_{\rm JT, 3D} = -\lambda_Q \sqrt{Q_2^2+Q_3^2}$
which is obtained after minimizing JT electron-lattice coupling energy in manganites
with respect to the $e_g$ orbital state.~\cite{Ahn01}
In undoped manganites, $Q_3/Q_2$ is about 0.3 - 0.4
(Refs.~\onlinecite{Rodriguez-Carvajal98} and \onlinecite{Balagurov04}),
which allows an approximation
$E_{\rm JT, 3D} \approx -\lambda_Q |Q_2| [1 + (Q_3/Q_2)^2/2]$.
Further neglecting the small $(Q_3/Q_2)^2/2$ term, we get the 2D analog of the JT energy
$E_{\rm JT}$, with the 2D JT distortion $e_{\rm JT}$ corresponding to 3D JT distortion
$Q_2$ except for a normalization factor difference.

We expand $E_{\rm JT}$
and $E_{\rm tol}$ in the form of a  Taylor series in $e_{10}$, $e_{20}$, $e_{3s}$ and $e'_{3s}$.
Only the leading order energy terms being kept, our total energy expression per Mn ion
$E_{\rm tot}$ is given below.
\begin{eqnarray}
    E_{\rm tot}&=& E_{\rm har}+E_{\rm JT}+E_{\rm tol}+E_{\rm cubic}+E_{\rm quartic}, \\
    E_{\rm har}&=& \frac{1}{2} (a_1+a_2) (e_{10})^2 + \frac{1}{2}(4
    b_1) (e_{20})^2 \nonumber \\
    & &+\frac{1}{2} a_1 e_{3s}^2 + \frac{1}{2} a_2 (e'_{3s})^2, \\
    E_{\rm JT}&=&-\lambda |e_{3s} + e_{20} e'_{3s}|, \label{eq:EJT} \\
    E_{\rm tol}&=& \tilde{p} e_{10}, \\
    E_{\rm cubic}&=& \frac{1}{2} a_1 e_{10} (e'_{3s})^2, \\
    E_{\rm quartic}&=&\frac{1}{4} \frac{a_1}{2} (e'_{3s})^4,
\end{eqnarray}
where the relation
\begin{equation}
e_{\rm JT}\approx e_{3s}+e_{20}e'_{3s} \label{eq:eJT}
\end{equation}
is used for $E_{\rm JT}$.
The physical origin of the coupling between $e_{20}$ and $e'_{3s}$
is important for the
current study and is explained in more detail in Sect.~\ref{sec:appearance}.

\section{Estimation of parameters}\label{sec:estimation}
In this section, we present our estimate of the parameters.
We choose the Mn-Mn distance before the distortion, which is around $u$ = 4 {\AA}, as 1. Therefore,
$e_{10}$, $e_{20}$, $e_{3s}$, and $e'_{3s}$ are unitless,
and $a_1$, $b_1$, $a_2$, $b_2$, and $\lambda $ have the
unit of energy. The parameter $a_1$ can be estimated from the Mn-O bond stretching phonon
mode energy, which is about 70 meV from optical measurements.~\cite{Ahn01}
From $\hbar \sqrt{2 a_1/m_{\rm O}}$ = 70 meV with $m_{\rm O}$ the mass of the oxygen ion, we obtain
$a_1 \approx $ 150 eV. We estimate $b_1$ from the elastic modulus, $c_{44}$. From
Ref.~\onlinecite{Darling98}, $c_{44} \approx $ 55 - 60 GPa. The uniform shear mode $e_{20}$
corresponds to the conventional $e_{xy}/2$ (Ref.~\onlinecite{Ashcroft}).
Using the identity 1 GPa
{\AA}$^3$ = 6.3 meV, we find $b_1 \approx $ 20 - 25 eV. To estimate $b_2$, we use the
results~\cite{Mirgorodsky93} for ReO$_3$, which has no $RE/AK$ ion and, therefore, $a_2=0$
and the buckling of Re-O-Re bond depends only on $b_2$. According to the analysis in
Ref.~\onlinecite{Mirgorodsky93}, the oxygen oscillation along Re-O-Re direction has the
angular frequency $\omega_o^x$ = 905 cm$^{-1}$, whereas the oscillation perpendicular to
Re-O-Re direction has the angular frequency $\omega_o^y$ = 30 cm$^{-1}$, from which we
can estimate $b_2/a_1=(\omega_o^y/\omega_o^x)^2 /2 \approx 0.5 \times 10^{-3}$. We can expect
a similar order of magnitude for $b_2$ in manganites, order of $10^{-3} a_1$, for example,
0.2 eV, which is negligible compared to other parameter values and justifies neglecting
the terms with $b_2$ as mentioned above. Various probes, such as neutron or optical spectroscopy,
indicate the buckling mode frequency in manganites of about 35 - 50 meV (Ref.~\onlinecite{Zhang01}).
From the analysis of $(\pi,\pi)$ phonon mode for our model, we obtain the frequency of buckling mode
$\omega_{\rm bk}=\sqrt{(2a_2+4b_2)/m_{\rm O}}$. Therefore, we obtain $a_2 \approx$ 30 - 80 eV.
For the estimation of $\lambda$, we match the JT energy gain for our 2D model with that for the 3D model
to ensure that our 2D model represents the energy scale of the 3D materials correctly.
For our 2D model $\Delta E_{\rm JT} = - \lambda^2/(2 a_1)$. For the 3D model in Ref.~\onlinecite{Ahn01},
$\Delta E_{\rm JT} \approx -$0.29 eV, and therefore, we obtain $\lambda \approx$ 10 eV.

\section{Interplay between ${\rm\bf Mn-O-Mn}$ bond buckling and the Jahn-Teller distortions}

\subsection{Buckling instability without the Jahn-Teller term}
We find the condition for the buckling instability without the effect of the JT
energy term $E_{\rm JT}$. We take a perturbative approach rather than try to solve high
order polynomial equations. By minimizing $E_{\rm har}+E_{\rm tol}$, we obtain
\begin{equation}
    (e_{10})^{\min,*}=-\frac{\tilde{p}}{a_1+a_2},
\end{equation}
where the superscript * indicates that the JT term is not yet taken into consideration.
This isotropic compression of the MnO$_4$ motif renormalizes the coefficient of the
$(e'_{3s})^2$ term through the $E_{\rm cubic}$ term. From this, we obtain the
critical condition for the buckling instability,
\begin{eqnarray}
    \tilde{p}_c^*&=&\frac{a_2}{a_1} (a_1+a_2), \label{eq:tcwoJT} \\
    (e_{10})_c^{\min,*}&=&-\frac{a_2}{a_1}.
\end{eqnarray}
If $\tilde{p} > \tilde{p}_c^*$, Mn-O-Mn bond buckling occurs and
the quartic order term, $E_{\rm quartic}$,
should be considered for the equilibrium $e'_{3s}$,
\begin{eqnarray}
    |(e'_{3s})^{\min,*}|&=&\sqrt{\frac{2}{a_1+a_2}} \sqrt{\tilde{p}-\tilde{p}_c^*} \\
    &=&\sqrt{2} \sqrt{(e_{10})_c^{\min,*}-(e_{10})^{\min,*}}.
\end{eqnarray}
The minimized $E_{\rm tot}$ without the $E_{\rm JT}$ term is given by
\begin{equation}
    E^{\min,*}_{\rm tot}=-\frac{\tilde{p}^2}{2(a_1+a_2)}
    -\frac{a_1}{2} \left( \frac{\tilde{p}}{a_1+a_2} -\frac{a_2}{a_1} \right)^2.
\end{equation}

\subsection{Buckling instability with the Jahn-Teller term}
We now examine how the JT energy term $E_{\rm JT}$ alters the buckling instability.
From
$E_{\rm har}+E_{\rm tol}+E_{\rm JT}$, we obtain
\begin{eqnarray}
    (e_{10})^{\min}&=&-\frac{\tilde{p}}{a_1+a_2}, \\
    (e_{3s})^{\min}&=&\frac{\lambda}{a_1}, \label{eq:e3s}
\end{eqnarray}
where we consider the $(e_{3s})^{\min}>0$ case only. The buckling instability is found from
the second order terms in $e_{20}$ and $e'_{3s}$ in $E_{\rm tot}$:
\begin{equation}
    \frac{1}{2}(4 b_1) (e_{20})^2 +
    \frac{1}{2} [a_2+a_1 (e_{10})^{\min}] (e'_{3s})^2 - \lambda e_{20} e'_{3s},
\end{equation}
where we assumed $(e_{3s})^{\min} + e_{20} e'_{3s} > 0 $. From the condition
$4 b_1 [a_2+a_1 (e_{10})^{\min}] < \lambda^2$, we obtain the critical condition
\begin{equation}
    \tilde{p}_c=\frac{a_2}{a_1} (a_1+a_2)-\frac{\lambda^2}{4 b_1 a_1} (a_1+a_2)
\end{equation}
and the buckling distortion occurs for $\tilde{p} > \tilde{p}_c$. Comparing with
$\tilde{p}_c^*$ in Eq.~(\ref{eq:tcwoJT}), we find that the JT energy makes buckling
more likely. After this buckling instability, we should include the $E_{\rm quartic}$ term
to find the equilibrium result. For this, we first minimize $E_{\rm tot}$ with respect
to the shear distortion $e_{20}$ to obtain
\begin{equation}
    (e_{20})^{\min}= \frac{\lambda}{4 b_1} e'_{3s}. \label{eq:e20}
\end{equation}
Inserting this back, we get an energy expression for $E_{\rm tot}$ only in terms of $e'_{3s}$,
which gives the equilibrium buckling distortion and the minimum energy,
\begin{equation}
    (e'_{3s})^{\min}=\sqrt{\frac{2}{a_1+a_2}} \sqrt{\tilde{p} - \tilde{p}_c},
\end{equation}
\begin{eqnarray}
    E_{\rm tot}^{\min} &=& -\frac{\tilde{p}^2}{2(a_1+a_2)} - \frac{\lambda^2}{2 a_1} \nonumber \\
    && -\frac{a_1}{2} \left( \frac{\tilde{p}}{a_1+a_2} -\frac{a_2}{a_1}
        +\frac{\lambda^2} {4b_1 a_1} \right)^2. \label{eq:Etotmin0}
\end{eqnarray}
Therefore, the energy gain due to the JT energy term is given by
\begin{equation}
    \Delta E_{\rm JT} = -\frac{\lambda^2}{2 a_1}
    -\frac{(\tilde{p}-\tilde{p}^*_c) \lambda^2 }{4 (a_1+a_2) b_1}
\end{equation}
up to order $\lambda^2$. The second term corresponds to the part of $\Delta E_{\rm JT}$ which
depends on the size of rare earth ion, or $\tilde{p}$. This result shows that the small rare
earth ion, or large chemical pressure, stabilizes the JT distortion.

\section{Comparison with experiments}
We make comparisons between our model and experimental results.
In Sect.~\ref{sec:appearance}, we explain the simultaneous appearance of the uniform shear distortion
and the long range JT distortion observed in undoped manganites.~\cite{Rodriguez-Carvajal98}
In Sect.~\ref{sec:JT-ordering}, we estimate the changes in the
JT ordering temperature $T_{\rm JT}$ between LaMnO$_3$ and NdMnO$_3$,
and compare with experiments.
In Sect.~\ref{sec:ratios}, we calculate the ratios between different distortion modes
and compare with experimental data for LaMnO$_3$ and NdMnO$_3$.

\subsection{Appearance of uniform shear distortion below the Jahn-Teller ordering temperature}\label{sec:appearance}

\begin{figure}
    \leavevmode
    \epsfxsize8.6cm \epsfbox{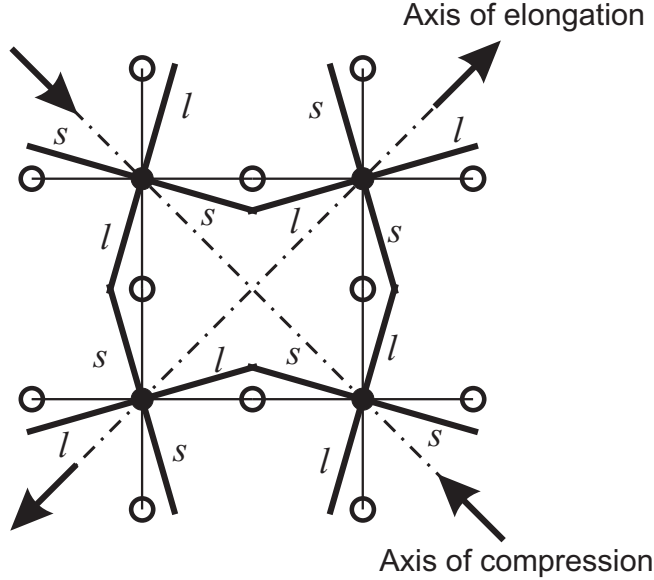}
    \caption{\label{fig:buckling}
    Superposition of $(\pi,\pi)$ buckling $e'_{3s}$ and uniform shear distortion $e_{20}$ effectively
    generates the extra $(\pi,\pi)$ JT distortion of Mn-O bond lengths, as indicated by
    the $s$ and $l$ for the shortened and elongated bonds, which is responsible for the
    $e_{20}e'_{3s}$ coupling within $e_{\rm JT}$ and the Jahn-Teller coupling $E_{\rm JT}$.
    In the $(\pi,\pi)$ JT ordered state,
    this adds up to the $e_{3s}$ deviatoric mode
    if $e_{3s}$ and $e_{20}e'_{3s}$ have the same sign.
    This mechanism is responsible for the appearance of the uniform shear distortion
    below the JT ordering temperature, as explained in Sect.\ref{sec:appearance}.
    The extra JT energy gain for the buckled lattice is responsible
    for the increase in $T_{\rm JT}$ in $RE$MnO$_3$ with small $RE$ ions, as explained in Sect.\ref{sec:JT-ordering}.
    }
\end{figure}

Experimental data in Refs.~\onlinecite{Rodriguez-Carvajal98} and~\onlinecite{Balagurov04}
show that the difference between the lattice constant $a$ and $b$
along the diagonal directions in the plane appears
simultaneously with the long range JT distortion
below $T_{\rm JT}$ for both LaMnO$_3$ and NdMnO$_3$.
This distortion corresponds to the uniform shear distortion in our model, related by
$e_{20}=(b-a)/(2\sqrt{2} u)$ with $u$ = 4 \AA.
We analyze the coupling between the JT distortion and the uniform shear distortion,
which is important for the stabilization of JT ordered state by the chemical pressure.
In our model, such coupling originates from
the term $e_{20} e'_{3s}$ in $e_{\rm JT}$ in Eq.~(\ref{eq:eJT}) or in $E_{\rm JT}$ in Eq.~(\ref{eq:EJT}),
which can be understood as follows.
We consider applying a positive $e_{20}$ shear distortion to the lattice,
as shown in Fig.~\ref{fig:buckling} by the axis of elongation and compression along 45$^{\circ}$ and
135$^{\circ}$, respectively. Such uniform shear distortion makes the Mn-O bond lengths either longer or shorter
depending on whether the direction of the bond is closer to the orientation of elongation
(45$^{\circ}$) or compression (135$^{\circ}$), except for the bonds with directions right between the two directions.
If the system {\it does not} have $(\pi,\pi)$ buckling, as shown in the {\it thin} solid
lines in Fig.~\ref{fig:buckling}, all Mn-O bonds make equal angles from the axis of
elogation/compression, and therefore $e_{20}$ shear distortion keeps all Mn-O bond lengths equal.
This implies that $e_{20}$ distortion alone does not contribute to the JT distortion or JT energy gain.
In contrast, if the system has a buckling distortion $e'_{3s}$  with a wave vector $\vec{k}=(\pi,\pi)$,
as shown in the {\it thick} solid lines in Fig.~\ref{fig:buckling}, the $e_{20}$ shear
distortion elongates Mn-O bonds marked with {\it l} and shortens Mn-O bonds marked with
{\it s}, depending on whether the bond direction is closer to
the axis of elongation or the axis of compression, which results in the JT
distortion $e_{\rm JT}$ with a wave vector $\vec{k}=(\pi,\pi)$. If this extra JT distortion
is in the same [opposite] phase as [to] the deviatoric $e_{3s}$ distortion, in other words, if $e_{20}e'_{3s}$
and $e_{3s}$ have the same [opposite] sign, this extra JT distortion increases [decreases]
the net JT distortion,
which explains the expression of $E_{\rm JT}$ in Eq.~(\ref{eq:EJT}) or $e_{\rm JT}$ in Eq.~(\ref{eq:eJT}).
We emphasize here that the extra JT energy gain occurs only when the $e_2$, $e_3$
and $e'_3$ distortions are in the right phase with respect to each other.
Experiments~\cite{Rodriguez-Carvajal98} show that the $(\pi,\pi)$ Mn-O-Mn bond buckling persists
even above $T_{\rm JT}$ without much change in size.
However, above $T_{\rm JT}$, the coherent
$e_{3}$ distortion does not exist, and therefore the extra JT distortion due to the
uniform $e_{20}$ distortion in the presence of staggered buckling distortion would increase
the JT energy gain in some regions and decrease the JT energy gain in other regions,
and does not change the net JT energy. In other words, the energy gain due to the cooperative
effect between $e_3$, $e'_3$, and $e_2$ does not exist at $T > T_{\rm JT}$.
We therefore expect
that the $e_{20}$ mode does not exist above $T_{\rm JT}$ and appears simultaneously with the long range
JT ordering, consistent with the experimental results.

\subsection{Jahn-Teller ordering temperature and its variation between LaMnO$_3$ and NdMnO$_3$}\label{sec:JT-ordering}
It is reported~\cite{Kimura03}
that $T_{\rm JT}$ changes from 750 K for LaMnO$_3$ to 1100 K for NdMnO$_3$
by about $\Delta T_{\rm JT}$ = 350 K.
We estimate $\Delta T_{\rm JT}$ from
our model to understand how such a drastic change of the JT ordering temperature
can occur by the increase in chemical pressure.

We rewrite $E^{\min}_{\rm tot}$ in Eq.~(\ref{eq:Etotmin0})
for $\tilde{p} > \tilde{p}_c$ as follows.
\begin{equation}
    E_{\rm tot}^{\min}=-\frac{\tilde{p}^2}{2(a_1+a_2)} - \frac{\lambda^2}{2 a_1}
    - \frac{a_1 ( \tilde{p} -\tilde{p}_c )^2}{2 (a_1+a_2)^2} ,
\end{equation}
where
\begin{eqnarray}
    \tilde{p}_c &=& \tilde{p}_{c}^*-\delta \tilde{p}_c, \\
    \delta \tilde{p}_c &=& \frac{\lambda^2}{4 b_1 a_1} (a_1+a_2).
\end{eqnarray}
Since $\delta \tilde{p}_c$, the change in the critical chemical pressure due to the
$E_{\rm JT}$ term, is small relative to $\tilde{p}-\tilde{p}_c^*$,
with $\delta\tilde{p}_c/(\tilde{p}-\tilde{p}_c^*)\approx$ 0.3 for parameter values in Sect.~\ref{sec:estimation},
we keep the terms up to linear in $\delta \tilde{p}_c$
only and rewrite according to the origin of each term as follows.
\begin{eqnarray}
    E_{\rm tot}^{\min} &\approx& E^{\min}_{\rm comp}+E^{\min}_{\rm JT}+E^{\min}_{\rm bk}+E^{\min}_{\rm bk,JT,sh}, \label{eq:Etotmin} \\
    E^{\min}_{\rm comp} &=& -\frac{1}{2} \frac{\tilde{p}^2}{a_1+a_2}, \\
    E^{\min}_{\rm JT} &=& -\frac{1}{2} \frac{\lambda^2}{a_1}, \\
    E^{\min}_{\rm bk} &=& -\frac{1}{2} \frac{a_1}{(a_1+a_2)^2} \left( \tilde{p} -\tilde{p}_{c}^* \right)^2, \\
    E^{\min}_{\rm bk,JT,sh} &=& -\frac{a_1}{(a_1+a_2)^2} \left( \tilde{p} -\tilde{p}_{c}^* \right) \delta \tilde{p}_c, \nonumber \\
    &=&-\frac{\lambda^2}{4 b_1 (a_1 + a_2)} (\tilde{p}-\tilde{p}_c^*).
\end{eqnarray}
The first three terms, $E^{\min}_{\rm comp}$, $E^{\min}_{\rm JT}$, and $E^{\min}_{\rm bk}$,
represent the energy terms purely due to compression, JT distortion, and buckling, respectively.
The fourth term is the energy due to the coherent buckling, JT and shear distortions,
indicated by its dependence on $\tilde{p}-\tilde{p}_c^*$, $\lambda$ and $b_1$,
which gives extra stability to the JT ordering due to the chemical pressure.

To estimate $T_{\rm JT}$, we consider a high temperature state with random JT distortions,
for which the energy can be written in a similar way as Eq.~(\ref{eq:Etotmin}) except for
the absence of the fourth term
due to the lack of coherence among distortions as explained in Sect.~\ref{sec:appearance},
\begin{equation}
    E_{\rm tot}^{\rm ran} = E^{\rm ran}_{\rm comp} + E^{\rm ran}_{\rm JT} + E^{\rm ran}_{\rm bk}.
\end{equation}
We expect $E^{\rm ran}_{\rm comp} \approx E^{\min}_{\rm comp}$ and
$E^{\rm ran}_{\rm bk} \approx E^{\min}_{\rm bk}$ since the unit cell volume and buckling
angle do not change very much as the temperature crosses $T_{\rm JT}$ (Ref.~\onlinecite{Rodriguez-Carvajal98}).
Therefore, the energy difference between JT ordered and JT disordered state is
\begin{equation}
    E^{\rm ran}_{\rm tot}-E^{\min}_{\rm tot} \approx E^{\rm ran}_{\rm JT}
        -E^{\min}_{\rm JT}-E^{\min}_{\rm bk,JT,sh}. \label{eq:TJT}
\end{equation}

We first verify that our model gives the correct order of magnitude of $T_{\rm JT}$ itself.
An order of magnitude estimate for $T_{\rm JT}$ can be made from the energy difference between two different
JT ordered states, one the most favored state and the other relatively unfavored state. The most
favored state is that with the JT distortion of $\vec{k}=(\pi,\pi)$ considered
so far in this paper and has the JT energy of $E_{\rm JT}^{\min} = - \lambda^2/(2a_1)$.
We choose a state with the same size of JT distortion $e_3$ but with a wave vector $\vec{k}=(0,0)$
as a relatively unfavored state, with energy $E_{\rm JT}^{\rm unif} = - \lambda^2/[2(a_1+a_2)]$.
Using the estimated parameter values, $a_1$ = 150 eV, $a_2$ = 30 - 80 eV, $\lambda $ = 10 eV,
we obtain $E_{\rm JT}^{\rm unif}-E_{\rm JT}^{\rm min} \approx$ 600 - 1300 K, which has the same
order of magnitude as the experimentally observed $T_{\rm JT}$ in the range of 750 K - 1100 K.

For the change in $T_{\rm JT}$ between LaMnO$_3$ and NdMnO$_3$, the only term in Eq.~(\ref{eq:TJT})
which changes with the $RE$ ion size is $-E^{\min}_{\rm bk,JT,shear}$.
Therefore, the JT ordering
temperature variation between LaMnO$_3$ and NdMnO$_3$ can be related to
$-E^{\min}_{\rm bk,JT,sh}({\rm NdMnO_3}) + E^{\min}_{\rm bk,JT,sh}({\rm LaMnO_3})$
within a factor of the order of one.
We express $E_{\rm bk,JT,sh}^{\min}$ in terms of $(e'_{3s})^{\rm min}$,
\begin{equation}
    E_{\rm bk,JT,sh}^{\min} = -\frac{1}{2} \frac{\lambda^2}{4 b_1} [(e'_{3s})^{\min}]^2.
\end{equation}
According to the experimental data,~\cite{Kimura03,Rodriguez-Carvajal98,Balagurov04}
the Mn-O-Mn bond angle is 155$^{\circ}$
for LaMnO$_3$ and 150$^{\circ}$ for NdMnO$_3$, which corresponds to $(e'_{3s})^{\min}$
of about 0.22 and 0.27, respectively. These distortions, along with
parameter values $\lambda$ = 10 eV and $b_1$ = 20 -25 eV,
result in
$-E^{\min}_{\rm bk,JT,sh}({\rm NdMnO_3}) + E^{\min}_{\rm bk,JT,sh}({\rm LaMnO_3}) \approx $
12 - 16 meV $\approx$ 140 - 190 K. From a classical Monte Carlo simulation for the double-well
potential model in Ref.~\onlinecite{Ahn03}, we find that the structural ordering temperature
is about twice the energy difference between the distorted ground state and
undistorted high energy state.~\cite{MC}
Although such a relation would depend on the details of the model,
if we assume a similar situation in the current model, the JT
ordering temperature variation can be estimated as twice the energy difference, therefore,
$T{\rm_{JT}(NdMnO_3)} - T{\rm_{JT}(LaMnO_3)} \approx
2 \times [-E^{\min}_{\rm bk,JT,sh}({\rm NdMnO_3}) + E^{\min}_{\rm bk,JT,sh}({\rm LaMnO_3})]$ = 280 - 380 K,
which agrees well with the experimental change in $T_{\rm JT}$, 350 K.

This agreement shows that indeed the JT ordered state is more stabilized
when the buckling increases
for smaller rare earth ions for undoped compounds.
The relatively large increase in the JT ordering
temperature, both in theory and experimental data, shows that the interplay between the rare earth
ion size and the JT distortion is significant, and should be taken into account to explain
the well-known temperature-tolerance factor phase diagram of both undoped and doped perovskite manganites.

\subsection{Relation between shear, buckling, and deviatoric distortion}\label{sec:ratios}
Equations~(\ref{eq:e3s}) and~(\ref{eq:e20}) imply that the following quantities
remain constant regardless of the variation in chemical pressure:
\begin{eqnarray}
(e_{3s})^{\min}&=& \frac{\lambda}{a_1}, \\
\frac{(e_{20})^{\min}}{(e'_{3s})^{\min}} &=& \frac{\lambda}{4 b_1}, \\
\frac{(e_{20})^{\min}}{(e_{3s})^{\min} (e'_{3s})^{\min}} &=& \frac{a_1}{4 b_1}.
\end{eqnarray}
We calculate these quantities from the experimental data for LaMnO$_3$ and NdMnO$_3$,
and present the results in Table~\ref{table:parameters}, in which we also show the relation
between the distortion variables in our model and experimental parameters
and the estimate of $T_{\rm JT}$ and $\Delta T_{\rm JT}$ obtained in Sect.~\ref{sec:JT-ordering}.
The results show that
$(e_{3s})^{\min}$, $(e_{20})^{\min}/(e'_{3s})^{\min}$, and
$(e_{20})^{\min}/[(e_{3s})^{\min} (e'_{3s})^{\min}]$
remain constant within 2 \%, 10 \%, and 7 \%, respectively,
in spite of 32 \% and 19 \% changes in $(e_{20})^{\min}$ and $(e'_{3s})^{\min}$.
These values also agree well with theoretical estimates obtained from the parameters
in Sect.\ref{sec:estimation}.
The results underscore the strong coupling between these distortions, in particular,
the important role played by the uniform shear distortion in connecting the Jahn-Teller
and buckling distortions, an aspect neglected in the literature so far.

\begin{table*}[t]
\caption{\label{table:parameters} Parameters from experimental data and
comparison with theoretical estimates. Experimental data for the lattice constants,
bond lengths, and bond angles for LaMnO$_3$ and
NdMnO$_3$ are from Ref.~\onlinecite{Rodriguez-Carvajal98} and \onlinecite{Balagurov04}, respectively,
measured at room temperature.
}
\begin{tabular}{l|ll|c}
\hline \hline
    Parameters      & LaMnO$_3$         & NdMnO$_3$         & Theoretical estimates \\
\hline
 Lattice constant, $a$                & 5.54 \AA          & 5.414 \AA         &    \\
 Lattice constant, $b$                & 5.75 \AA          & 5.731 \AA         &    \\
 Long Mn-O bond length within $ab$ plane, $l$          & 2.718 \AA         & 2.20 \AA          &    \\
 Short Mn-O bond length within $ab$ plane, $s$          & 1.907 \AA         & 1.90 \AA          &    \\
 Mn-O-Mn bond angle within $ab$ plane, $\varphi$ & 155.1$^{\circ}$  & 149.8$^{\circ}$  &  \\
\hline
Mn-Mn distance with $e_{10}$ only, $(a+b)/(2\sqrt{2})$ & 3.995 \AA & 3.94 \AA & Compressed from $u \approx$ 4 \AA  \\
\hline
$e_{20}=(b-a)/(2\sqrt{2}u)$ & 0.019 & 0.028 &   \\
$e'_{3s}=(\pi-\varphi)/2$ & 0.22 & 0.27 &  \\
$e_{20}/e'_{3s}$ & 0.09 & 0.10 & $\lambda/(4b_1) = 0.10 - 0.13$ \\
$e_{\rm JT}=(l-s)/u$ & 0.068 & 0.075 &  \\
$e_{3s}\approx e_{\rm JT} - e_{20}e'_{3s}$ & 0.064 & 0.067 & $\lambda/a_1$ = 0.067 \\
$e_{20}/(e_{3s}e'_{3s})$ & 1.4 & 1.5 & $a_1/(4b_1) = 1.5 - 1.9$ \\
\hline
$T_{\rm JT}$ & 750 K & 1100 K & $E_{\rm JT}^{\rm unif} - E_{\rm JT}^{\rm min}  = 600 - 1300$ K \\
$\Delta T_{\rm JT}$ & \multicolumn{2}{c|}{350 K} & $-2\Delta E^{\min}_{\rm bk,JT,sh} = 280 - 380$ K \\
\hline \hline
\end{tabular}
\end{table*}

\section{Conclusion}
From the analysis of a Keating energy expression expanded in terms of the atomic-scale symmetry-modes,
we find that the effect of small rare earth ion size, known as chemical pressure effect,
is significant in stabilizing the long range Jahn-Teller distortion in undoped perovskite manganites.
We obtain  good agreement with the experimental data on the Jahn-Teller ordering temperature and the
substantial increase of the Jahn-Teller ordering temperature from LaMnO$_3$ to NdMnO$_3$.
We propose that similar effects need to be considered
to understand the phase diagram for doped perovskite manganites.
We also explain the appearance of the uniform shear distortion below the Jahn-Teller ordering temperature
in terms of the coupling between coherent shear, buckling, and deviatoric distortions within the Jahn-Teller energy.
Moreover, we estimate the ratio between these distortions at low temperature,
and find good agreement with experimental data for LaMnO$_3$ and NdMnO$_3$,
which confirms the coupling  proposed between them in our model.

This work was supported by US DOE/LANL Award No. DE-AC52-06NA25396/170590-1 (T.F.S., K.H.A.),
ANL XSD Visitor Program (K.H.A.),
NJIT (T.F.S., K.H.A.),
US DOE LANL LDRD (T.L., A.S., A.R.B.),
and DOE FWP 70069 (P.B.L.).



\begin{thebibliography}{99}
\bibitem{Norm}
    The normalization factor is chosen in such a way that, for example,
    $e_3 = 2 d_{\rm O}$ if each O ion around Mn ion is displaced by $d_{\rm O}$
    in the way shown in Fig.~\ref{fig:modes}.
\bibitem{MC}
    We carry out classical Monte Carlo simulations with the same energy expression and
    parameter values used for the results in Fig.~3 in Ref.~\onlinecite{Ahn03}.
    We find that the structural phase transition temperature is about 0.15, which is about
    twice the depth of the potential well 0.08.
\end{thebibliography}
\end{document}